\documentclass[twocolumn,a4paper,
superscriptaddress,floatfix,twoside]{revtex4}

\usepackage{bm}
\usepackage{epsfig}
\usepackage{graphics}
\usepackage{natbib}

\usepackage{fancyh}
\usepackage[l]{floatflt}
\usepackage{epsfig}
\usepackage{amssymb}
\usepackage{latexsym}
\usepackage{times}
\usepackage{slashed}
\usepackage{upgreek}
\usepackage{amsmath}
\usepackage{amsfonts}
\usepackage{amsbsy}
\usepackage{amscd}
\usepackage{bbm}

\newcommand{\eec}{\end{center}}
\newcommand{\bec}{\begin{center}}

\newcommand{\eem}{\end{matrix}}
\newcommand{\bem}{\begin{matrix}}
\newcommand{\eeq}{\end{equation}}
\newcommand{\beq}{\begin{equation}}
\newcommand{\ba}{\begin{array}}
\newcommand{\ea}{\end{array}}
\newcommand{\bea}{\begin{eqnarray}}
\newcommand{\eea}{\end{eqnarray}}
\newcommand{\baq}{\begin{eqnarray}}
\newcommand{\eaq}{\end{eqnarray}}
\newcommand{\beqs}{\begin{subequations}}
\newcommand{\eeqs}{\end{subequations}}

\newcommand\eqs[2]{Eqs.~(\ref{#1}) and (\ref{#2})}

\newcommand{\ftn}{\footnotesize}

\newcommand{\TeV}{{\mbox{\rm TeV}}}

\newcommand{\GeV}{{\mbox{\rm GeV}}}
\newcommand{\eV}{{\mbox{\rm eV}}}
\newcommand{\sFref}[2]{Fig.~\ref{#1}-{\small \sf ({#2})}}

\newcommand{\etal}{{\it et al.\/}}

\def\to{\rightarrow}

\def\lf{\left(}
\def\rg{\right)}
\newcommand\vev[1]{\langle {#1} \rangle}
\newcommand{\Gr}{\ensuremath{\widetilde{G}}}
\newcommand{\Yb}{\ensuremath{Y_{B}}}

\newcommand{\Nhi}{\ensuremath{N_{\rm HI*}}}

\newcommand{\Vhi}{\ensuremath{V_{\rm HI}}}

\newcommand{\Whi}{\ensuremath{W_{\rm HI}}}

\def\ve{\varepsilon}
\newcommand{\Vhio}{\ensuremath{V_{\rm HI0}}}

\newcommand{\mP}{\ensuremath{m_{\rm P}}}
\newcommand{\Mpq}{\ensuremath{M}}
\newcommand{\Mgut}{\ensuremath{M_{\rm GUT}}}
\newcommand{\Ggut}{\ensuremath{G_{B-L}}}

\newcommand{\aS}{\ensuremath{{\rm a}_S}}

\newcommand{\msn}{\ensuremath{m_{\rm I}}}
\newcommand{\mgr}{\ensuremath{m_{3/2}}}

\newcommand{\hu}{{\ensuremath{H_u}}}

\newcommand{\ns}{\ensuremath{n_{\rm s}}}
\newcommand{\as}{\ensuremath{\alpha_{\rm s}}}
\newcommand{\sni}{\ensuremath{\nu^c_i}}

\newcommand{\wrhn[1]}{\ensuremath{\nu^c_{#1}}}

\newcommand{\Dex}{\ensuremath{\Delta_{\rm c*}}}
\newcommand{\Dmax}{\ensuremath{\Delta_{\rm max*}}}

\newcommand{\br}{\ensuremath{{\sf Br}}}

\newcommand{\Trh}{\ensuremath{T_{\rm rh}}}
\newcommand{\sg}{\ensuremath{\sigma}}
\newcommand{\sgex}{\ensuremath{\sigma_*}}
\newcommand{\sgc}{\ensuremath{\sigma_{\rm c}}}
\newcommand{\sgmax}{\ensuremath{\sigma_{\rm max}}}

\newcommand{\ld}{\ensuremath{\lambda}}

\newcommand{\kp}{\ensuremath{\kappa}}

\newcommand{\hepph}[1]{{\ftn\tt hep-ph/#1}}

\newcommand{\astroph}[1]{{\ftn\tt astro-ph/#1}}
\newcommand{\arxiv}[1]{{\ftn\tt  arXiv:#1}}

\newcommand{\Eref}[1]{Eq.~(\ref{#1})}
\newcommand{\Sref}[1]{Sec.~\ref{#1}}
\newcommand{\Fref}[1]{Fig.~\ref{#1}}
\newcommand{\Tref}[1]{Table~\ref{#1}}
\newcommand{\cref}[1]{Ref.~\cite{#1}}

\def\FHI{FHI~}
\def\Ka{K\"{a}hler potential}

\newcommand{\bdhh}{{\ensuremath{\normalsize I{\kern-2.9pt H}}}}

\newcommand{\mrh[1]}{\ensuremath{M_{#1\nu^c}}}

\newcommand{\mD[1]}{\ensuremath{m_{#1\rm D}}}
\newcommand{\mn[1]}{\ensuremath{m_{#1\nu}}}
\newcommand{\Gm[1]}{\ensuremath{\Gamma_{#1}}}

\renewcommand{\refname}{{\bf\scshape References}}

\renewenvironment{subequations}{%
\refstepcounter{equation}%
\setcounter{parentequation}{\value{equation}}%
  \setcounter{equation}{0}
  \ignorespaces
}{%
  \setcounter{equation}{\value{parentequation}}%
  \ignorespacesafterend
}

\begin{document}


\title{\bf\scshape Update on Minimal Supersymmetric Hybrid Inflation in Light of PLANCK}

\author{\scshape Constantinos Pallis}
\affiliation{Department of Physics, University of Cyprus, P.O. Box
20537, Nicosia 1678, CYPRUS
\\  {\sl e-mail address: }{\ftn\tt cpallis@ucy.ac.cy}}
\author{\scshape  Qaisar Shafi}
\affiliation{ Bartol Research Institute, Department of Physics and
Astronomy, University of Delaware, Newark, DE 19716, USA\\  {\sl
e-mail address: }{\ftn\tt shafi@bartol.udel.edu}}


\begin{abstract}

\noindent {\ftn \bf\scshape Abstract:} The minimal supersymmetric
(or F-term) hybrid inflation is defined by a unique renormalizable
superpotential, fixed by a $U(1)$ R-symmetry, and it employs a
canonical K\"{a}hler potential. The inflationary potential takes
into account both radiative and supergravity corrections, as well
as an important soft supersymmetry breaking term, with a mass
coefficient in the range $(0.1 - 10)~\TeV$. The latter term
assists in obtaining a scalar spectral index $n_s$ close to 0.96,
as strongly suggested by the PLANCK and WMAP-9yr measurements. The
minimal model predicts that the tensor-to-scalar $r$ is extremely
tiny, of order $10^{-12}$, while the spectral index running,
 $|d\ns/d\ln k| \sim 10^{-4}$. If inflation is associated with
the breaking of a local $U(1)_{B-L}$ symmetry, the corresponding
symmetry breaking scale $M$ is $(0.7 - 1.6)\cdot10^{15}~\GeV$ with
$n_s \simeq 0.96$. This scenario is compatible with the bounds on
$M$ from cosmic strings, formed at the end of inflation from $B-L$
symmetry breaking. We briefly discuss non-thermal leptogenesis
which is readily implemented in this class of models.
\\ \\ {\scriptsize {\sf PACs numbers: 98.80.Cq, 12.60.Jv}
\hfill {\sl\bfseries Published in} {\sl Phys. Lett. B } {\bf 725},
327 (2013)

}

\end{abstract}\pagestyle{fancyplain}

\maketitle

\rhead[\fancyplain{}{ \bf \thepage}]{\fancyplain{}{\sl Update on
Minimal SUSY Hybrid Inflation in Light of PLANCK}}
\lhead[\fancyplain{}{\sl \leftmark}]{\fancyplain{}{\bf \thepage}}
\cfoot{}

\section{Introduction}

\emph{Supersymmetric} (SUSY) hybrid inflation based on F-terms,
also referred to as \emph{F-term hybrid inflation} (FHI), is one
of the simplest and well-motivated inflationary models
\cite{susyhybrid,hybrid}. It is tied to a renormalizable
superpotential uniquely determined by a global $U(1)$ R-symmetry,
does not require fine tuned parameters, and it is naturally
associated with the breaking of a local symmetry, such as
$G_{B-L}= G_{\rm MSSM}\times U(1)_{B-L}$ \cite{bl}, where ${G_{\rm
MSSM}}= SU(3)_{\rm C}\times SU(2)_{\rm L}\times U(1)_{Y}$ is the
gauge group of the \emph{Minimal Supersymmetric Standard Model}
(MSSM) or, $G_{\rm LR}=SU(2)_{\rm L} \times SU(2)_{\rm R} \times
U(1)_{B-L}$ \cite{dvali}, flipped $SU(5)$ \cite{flipped}, etc. As
shown in \cref{susyhybrid}, the addition of \emph{radiative
corrections} (RCs) to the tree level inflationary potential
predicts a scalar spectral index $n_s\simeq0.98$, and the
microwave temperature anisotropy $\Delta T/T$ is proportional to
$(M / \mP)^2$, where $M$ denotes the scale of the gauge symmetry
breaking. It turns out that $M$ usually is not far from
$\Mgut\simeq (2-3)\cdot10^{16}~{\rm GeV}$. Here $\mP = 2.4 \cdot
10^{18}~\GeV$ is the reduced Planck mass. A more complete
treatment \cite{sstad2}, which incorporates \emph{supergravity}
(SUGRA) corrections \cite{senoguz} with canonical (minimal) \Ka,
as well as an important soft SUSY breaking term \cite{sstad1}, can
yield lower $\ns$ values ($0.95 - 0.97$). Recall that the minimal
\Ka\ insures that the SUGRA corrections do not spoil the flatness
of the potential that is required to implement FHI -- reduction of
$\ns$ by invoking non-minimal \Ka s is analyzed in \cref{gpp,
rlarge, hinova}.

Insisting on the simplest realization of FHI -- and the one-step
inflationary paradigm, cf. \cref{mhi} -- we wish to emphasize here
that FHI is in good agreement, in a rather narrow but well-defined
range of its parameters, with the latest WMAP \cite{wmap} and
PLANCK \cite{plin} data pertaining to the $\Lambda$CDM framework.
To this end, SUGRA \cite{senoguz} and soft SUSY breaking
\cite{sstad1, sstad2} corrections are taken into account, in
addition to the well-known \cite{susyhybrid} RCs. The minimality
of the model is justified by the fact that FHI is implemented
within \emph{minimal supergravity} (mSUGRA) and within a minimal
extension of ${G_{\rm MSSM}}$, obtained by promoting the
pre-existing global $U(1)_{B-L}$ symmetry of MSSM to a local one.
As a consequence, three \emph{right-handed neutrinos}, $\sni$, are
necessary to cancel the anomalies. The presence of $\sni$ leads to
a natural explanation for the observed \cite{plcp} \emph{baryon
asymmetry of the universe} (BAU) via \emph{non-thermal
leptogenesis} (nTL) \cite{lept}, and the existence of tiny but
non-zero neutrino masses. As we show, this set-up is compatible
with the gravitino constraint \cite{gravitino, kohri} and the
current data \cite{valle,lisi} on the neutrino oscillation
parameters. It is worth mentioning that our scenario fits well
with the bound \cite{plcs} induced by the non-observation of the
cosmic strings, formed during the $B-L$ phase transition. Note
that strings may serve as a source \cite{strings} of a
controllable amount of non-gaussianity in the cosmic microwave
background anisotropy.

In the following discussion, we briefly review the minimal FHI and
present our updated results in \Sref{fhi}. We then consider nTL
using updated constraints from neutrino physics in \Sref{pfhi}.
Our conclusions are summarized in \Sref{con}.

\section{Minimal FHI Model}\label{fhi}

\paragraph{\sf\scshape \small General Set-up.} The minimal FHI is based on the superpotential
\beq \Whi = \kappa S\left(\bar \Phi\Phi-M^2\right),\label{Whi}\eeq
where $\bar{\Phi}$, $\Phi$ denote a pair of chiral superfields
oppositely charged under $U(1)_{B-L}$, $S$ is a $\Ggut$-singlet
chiral superfield, and the parameters $\kappa$ and $M$ are made
positive by field redefinitions. $\Whi$ is the most general
renormalizable superpotential consistent with a continuous
R-symmetry \cite{susyhybrid} under which $S\  \to\
e^{i\alpha}\,S,~\bar\Phi\Phi\ \to\ \bar\Phi\Phi,~W \to\
e^{i\alpha}\, W$. The SUSY potential, $V_{\rm SUSY}$, extracted
(see e.g. \cref{lectures, review}) from $W_{\rm HI}$ in
Eq.~(\ref{Whi}) includes F and D-term contributions.  Along the
direction $|\bar\Phi|=|\Phi|$, the latter contribution vanishes
whereas the former reads
\beq \label{VF} V_{\rm SUSY}=
\kappa^2\left((|\Phi|^2-M^2)^2+2|S|^2 |\Phi|^2\right). \eeq
The scalar components of the superfields are denoted by the same
symbols as the corresponding superfields. Restricting ourselves to
the D-flat direction, from $V_{\rm SUSY}$ in Eq.~(\ref{VF}) we
find that the SUSY vacuum lies at
\beq
\vev{S}=0\>\>\>\mbox{and}\>\>\>\left|\vev{\Phi}\right|=\left|\vev{\bar\Phi}\right|=M.
\label{vevs} \eeq
As a consequence, $\Whi$ leads to the spontaneous breaking of
$\Ggut$, to $G_{\rm MSSM}$ with SUSY unbroken.

The superpotential $\Whi$ also gives rise to FHI since, for values
of $|S| \gg M $, there exist a flat direction
\begin{equation} \label{V0}\bar{\Phi}={\Phi}=0 ~~
\mbox{with,}~~V_{\rm SUSY}\lf{\bar{\Phi}=\Phi}=0\rg \equiv V_{\rm HI0}=\kappa^2 M^4.
\end{equation}
Thus, $V_{\rm HI0}$ provides us with a constant potential energy
density which can be used to implement FHI.

\paragraph{\sf\scshape \small The Inflationary Potential.}
The inflationary potential of minimal FHI, to a good
approximation, can be written as
\beq\label{Vol} V_{\rm HI}=V_{\rm HI0}+V_{\rm HIc}+V_{\rm
HIS}+V_{\rm HIT},\eeq where, besides the dominant contribution
$V_{\rm HI0}$ in \Eref{V0}, $\Vhi$ includes the following
contributions:

$\bullet$ $V_{\rm HIc}$ represents the RCs to $V_{\rm HI}$
originating from a mass splitting in the $\Phi-\bar{\Phi}$
supermultiplets, caused by SUSY breaking along the inflationary
valley \cite{susyhybrid}:
\beqs\beq \label{Vcor}V_{\rm HIc}=
{\kappa^2 {\sf N}\over 32\pi^2}\Vhio\left(2 \ln {\kappa^2x M^2
\over Q^2} +f_{\rm rc}(x)\right),\eeq
where ${\sf N}=1$ is the dimensionality of the representations to
which $\bar{\Phi}$ and $\Phi$ belong, $Q$ is a renormalization
scale, $x=\sigma^2/2M^2$ with $\sigma=\sqrt{2} \vert S\vert$ being
the canonically normalized inflaton field, and
\beq f_{\rm rc}(x)=(x+1)^{2}\ln\lf1+{1/
x}\rg+(x-1)^{2}\ln\lf1-{1/x}\rg.\label{frc}\eeq\eeqs
$\bullet$ $V_{\rm HIS}$ is the SUGRA correction to $V_{\rm HI}$ \cite{senoguz,sstad1}:
\begin{equation} \label{Vsugra}  V_{\rm HIS}=V_{\rm HI0}
{\sigma^4/8m^4_{\rm P}}, \end{equation}
where we employ the canonical K\"{a}hler potential $K =
|S|^2+|\Phi|^2+|\bar\Phi|^2$ working within mSUGRA.

$\bullet$ $V_{\rm HIT}$ is the most important contribution to
$V_{\rm HI}$ from the soft SUSY effects \cite{sstad1, sstad2}
parameterized as follows:
\begin{equation} \label{Vtad}  V_{\rm HIT}=-{\rm a}_S\,\sigma
\sqrt{V_{\rm HI0}/2},
\end{equation}
where \cite{dvali,sstad2} ${\rm
a}_S=2|2-A|\mgr\cos\lf\theta_S+\theta_{(2-A)}\rg$ is the tadpole
parameter which takes values comparable to $\mgr \sim
(0.1-10)~{\rm TeV}$, the gravitino, $\Gr$, mass. The soft SUSY
breaking mass$^{2}$ term for $S$, with mass $\sim \mgr$, is
negligible \cite{rlarge} for FHI. Also, $A$ is the dimensionless
trilinear coupling, of order unity, associated with the first term
of $\Whi$ in \Eref{Whi}. Imposing the condition
$\theta_S+\theta_{(2-A)}=0~{\sf\small mod}~2\pi$, $V_{\rm HI}$ is
minimized \emph{with respect to} (w.r.t.) the phases $\theta_S$
and $\theta_{(2-A)}$ of $S$ and $(2-A)$ respectively. We further
assume that $\theta_S$ remains constant during FHI.

\paragraph{\sf\scshape\small The Inflationary Observables -- Requirements.} Under
the assumptions that (i) the curvature perturbation generated by
$\sigma$ is solely responsible for the one that is observed, and
(ii) FHI is followed in turn by a decaying-particle, radiation and
matter domination, the parameters of our model can be restricted
by requiring that:

$\bullet$ The number of e-foldings $\Nhi$ that the scale
$k_*=0.05/{\rm Mpc}$ undergoes during FHI leads to a solution of
the horizon and flatness problems of standard big bang cosmology.
Employing standard methods \cite{hinova, plin, review}, we can
derive the relevant condition:
\begin{equation}  \label{Nhi}
\Nhi \equiv \int_{\sigma_{\rm f}}^{\sigma_{*}}\, \frac{d\sigma}{m^2_{\rm P}}\:
\frac{V_{\rm HI}}{V'_{\rm HI}}\simeq19.4+{2\over
3}\ln{V^{1/4}_{\rm HI0}\over{1~{\rm GeV}}}+ {1\over3}\ln {T_{\rm
rh}\over{1~{\rm GeV}}},
\end{equation}
where $T_{\rm rh}$ is the reheat temperature after FHI, the prime
denotes derivation w.r.t. $\sigma$, $\sigma_{*}$ is the value of
$\sigma$ when $k_*$ crossed outside the horizon of FHI, and
$\sigma_{\rm f}$ is the value of $\sigma$ at the end of FHI. This
coincides with either the critical point $\sigma_{\rm
c}=\sqrt{2}M$ appearing in the particle spectrum of
$\Phi-\bar\Phi$ system during FHI -- see \Eref{frc} -- or the
value for which one of the slow-roll parameters \cite{review}
\beq \label{slow} \epsilon\simeq{m^2_{\rm P}}\left({V'_{\rm
HI}}/{V_{\rm HI}}\right)^2/2~~\mbox{and}~~\eta\simeq m^2_{\rm
P}~{V''_{\rm HI}}/{V_{\rm HI}} \eeq
exceeds unity. In our scheme, we exclusively find $\sigma_{\rm
f}=\sigma_{\rm c}$. Since the resulting $\kp$ values are sizably
larger than $(M/\mP)^2$ -- see next section -- we do not expect
the production of extra e-foldings during the waterfall regime,
which in our case turns out to be nearly instantaneous -- cf.
\cref{bjorn}.

$\bullet$ The amplitude, $A_{\rm s}$, of the power spectrum of the
curvature perturbation, which is generated during FHI and
calculated at $k_{*}$ as a function of $\sg_*$, is consistent with
the data \cite{wmap, plin}, i.e.
\begin{equation} \label{Prob}
A_{\rm s}^{1/2}= \frac{1}{2\sqrt{3}\, \pi m^3_{\rm P}}\;
\left.\frac{V_{\rm HI}^{3/2}(\sigma_*)}{|V'_{\rm
HI}(\sigma_*)|}\right.\simeq\: 4.685\cdot 10^{-5}.
\end{equation}

$\bullet$ The (scalar) spectral index $n_{\rm s}$, its running,
${d\ns}/{d\ln k} \equiv \alpha_{\rm s}$, and the scalar-to-tensor
ratio, $r$,  which are given by
\beqs\bea \label{nS} && n_{\rm s}=1-6\epsilon_*\ +\ 2\eta_*,~~~~~~\\
&& \label{aS} \alpha_{\rm s}={2}\left(4\eta_*^2-(n_{\rm
s}-1)^2\right)/3-2\xi_*~~\mbox{and}~~ r=16\epsilon_*, ~~~~~~~
\eea\eeqs
where $\xi\simeq m_{\rm P}^4~V'_{\rm HI} V'''_{\rm HI}/V^2_{\rm
HI}$ and all variables with the subscript $*$ are evaluated at
$\sigma=\sigma_{*}$, should be in agreement with the following
values \cite{wmap, plin} based on the $\Lambda$CDM model:
\beqs\bea\label{nswmap} &&
\ns=0.9603\pm0.014~\Rightarrow~0.946\lesssim n_{\rm s}
\lesssim 0.975,~~~~~~\\
&&\label{obs3}\as=-0.0134\pm0.018,~~\mbox{and}~~
r<0.11, \label{obs4}\eea\eeqs
at 95$\%$ \emph{confidence level} (c.l.).

$\bullet$ The tension $\mu_{\rm cs}$ of the $B-L$ cosmic strings
produced at the end of FHI respects the bound \cite{plcs} -- cf.
\cref{jp,gmb,buch}:
\begin{equation} \label{mucs} \mu_{\rm cs} \approx
9.6\pi M^2/\ln(2/\beta)\lesssim 8\cdot10^{-6}\mP^2.\end{equation}
Here, we adapt to our set-up the results of the simulations for
the abelian Higgs model  following \cref{markjones},
$\beta=\kappa^2/8g^2\leq10^{-2}$, with $g\simeq0.7$ being the
gauge coupling constant close to $\Mgut$. Note that the presence
of strings does not anymore \cite{mark} allow $n_s$ closer to
unity.

\begin{figure*}[!t]
\centering
\includegraphics[width=60mm,angle=-90]{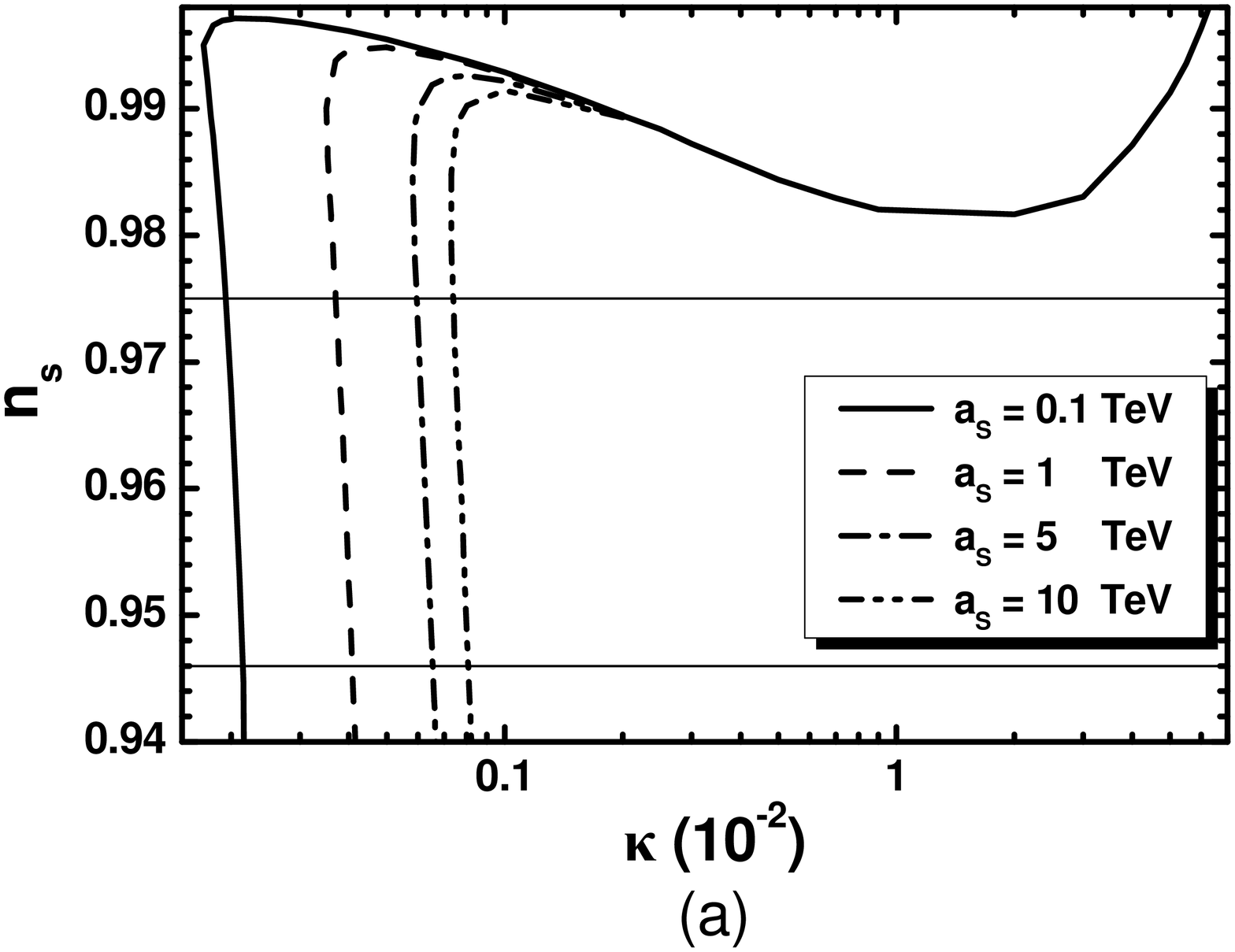}
\includegraphics[width=60mm,angle=-90]{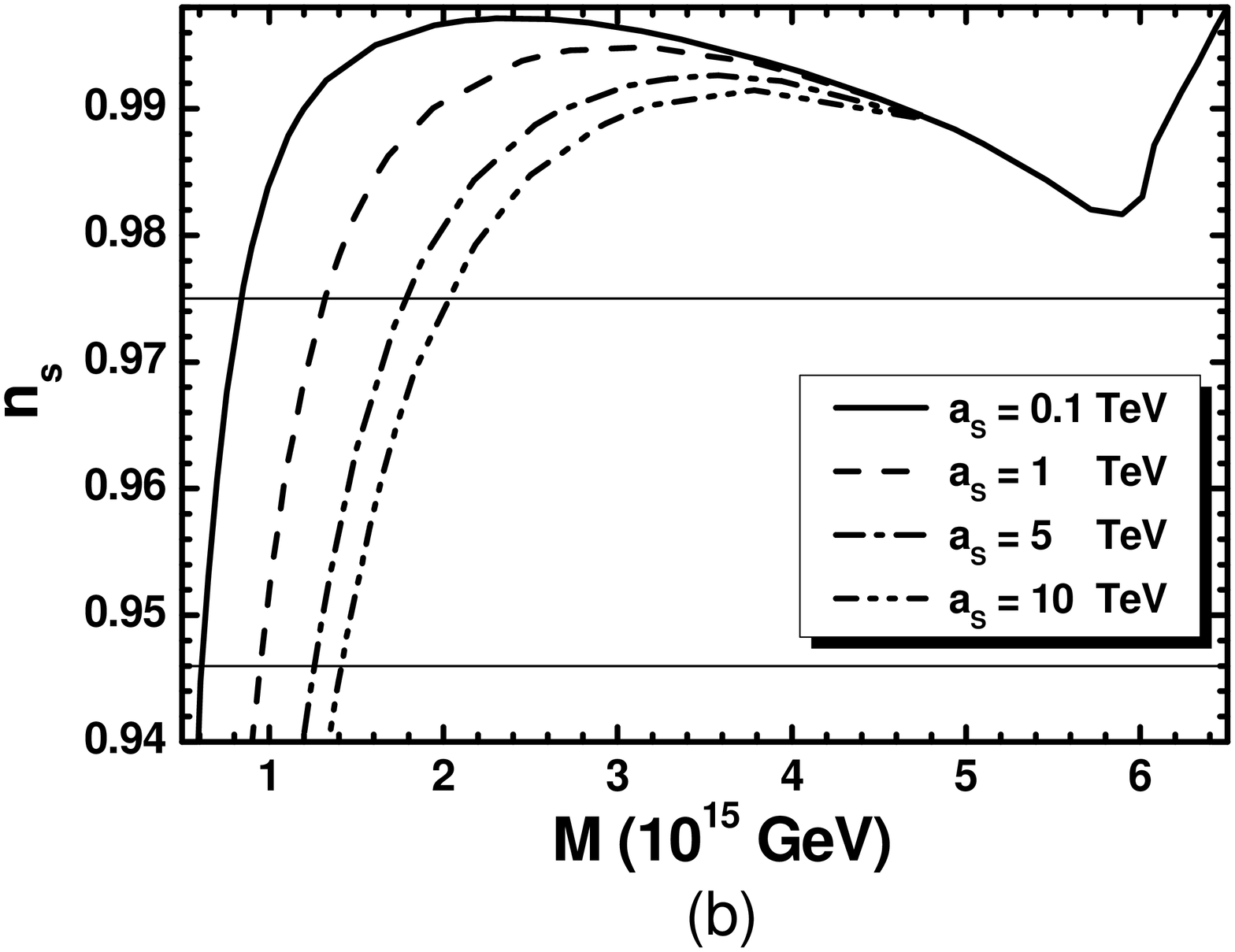}
\caption{\label{fig1}\sl $n_{\rm s}$ versus $\kappa$ (a), and
$n_{\rm s}$ versus $M$ (b) for $\aS=0.1~\TeV$ (solid lines),
$\aS=1~{\rm TeV}$ (dashed lines), $\aS=5~{\rm TeV}$ (dot-dashed
lines) and $\aS=10~{\rm TeV}$ (double dot-dashed lines). The two
horizontal lines are based on Eq.~(\ref{nswmap}) }
\end{figure*}
\begin{figure*}[!t]
\centering
\includegraphics[width=60mm,angle=-90]{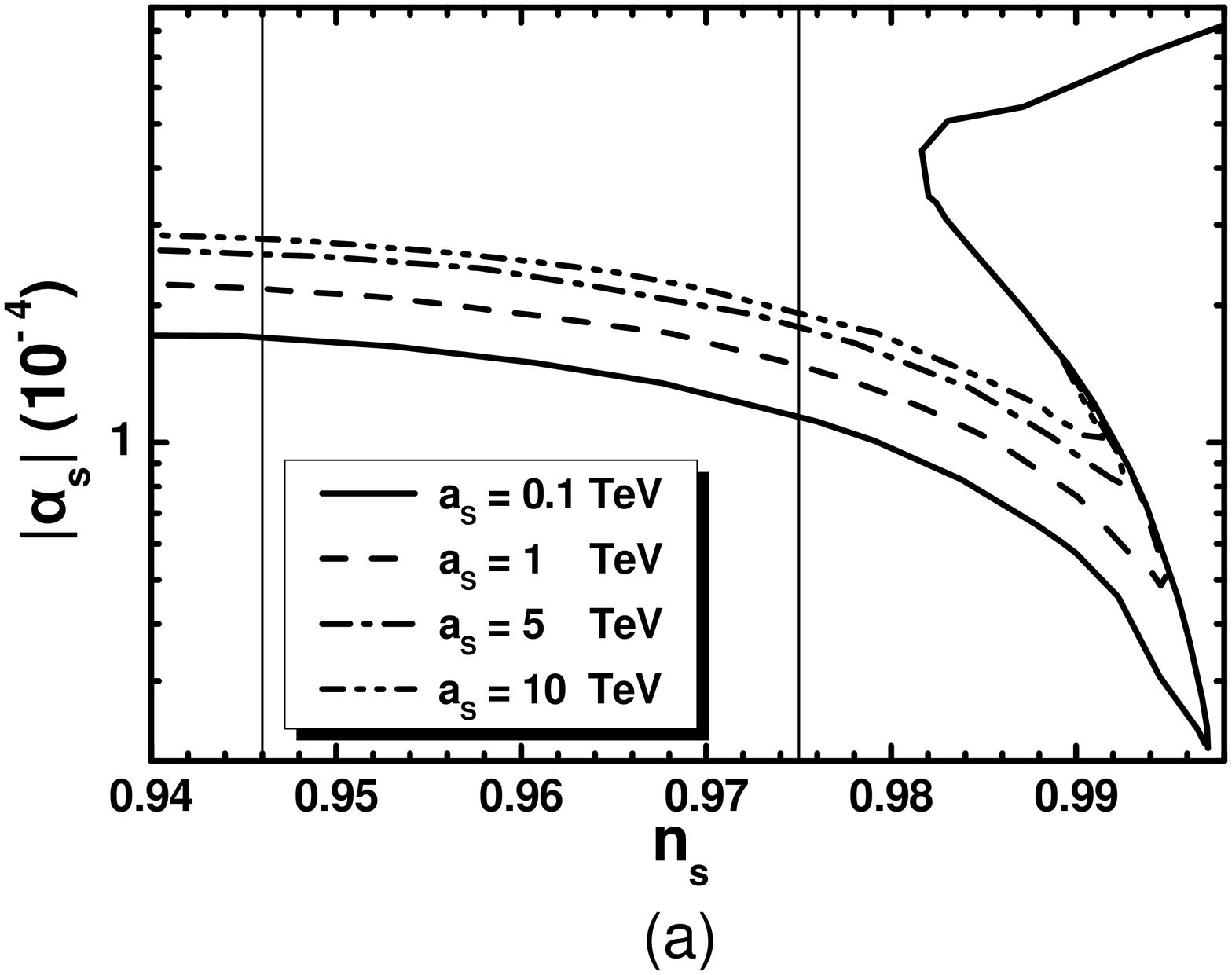}
\includegraphics[width=60mm,angle=-90]{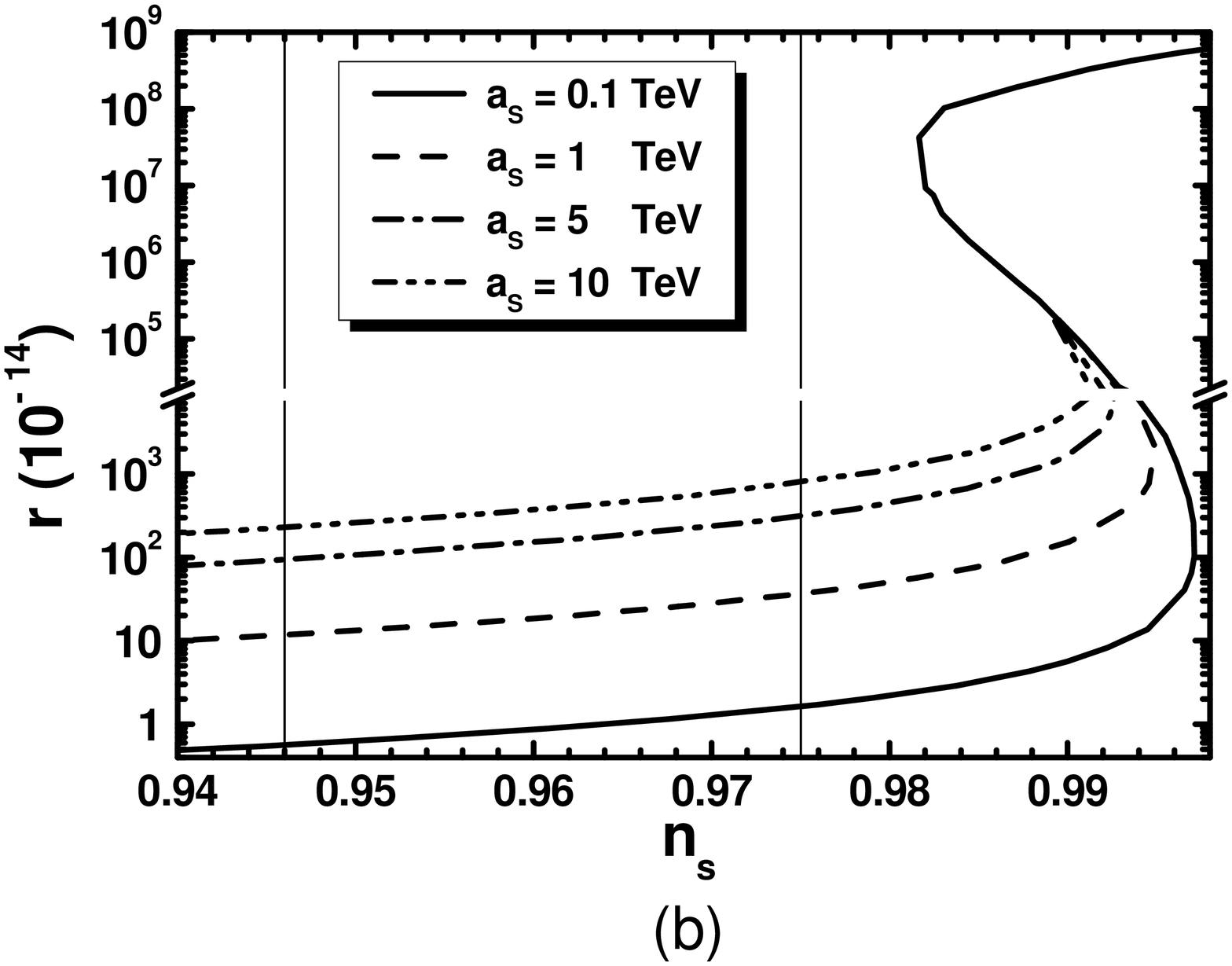}
\caption{\label{arns}\sl $|\as|$ versus $n_{\rm s}$ (a), and $r$
versus $\ns$(b) respectively. Vertical lines arise from
\Eref{nswmap}.}
\end{figure*}

\paragraph{\sf\scshape\small Results.} The investigation of our model depends on the
parameters:
$$ \kappa,~M, ~\aS,~\Trh,~~\mbox{and}~~\sigma_*\ .$$
In our computation, we use as input parameters $\aS$ and $\kappa$,
and fix $T_{\rm rh}\simeq5\cdot 10^{8}~\GeV$, as suggested by our
results in \Sref{pfhi}. Variation of $\Trh$ over $1-2$ orders of
magnitude is not expected to significantly alter our findings --
see \Eref{Nhi}. We then restrict $M$ and $\sigma_*$ so that
Eqs.~(\ref{Nhi}) and (\ref{Prob}) are fulfilled. Using
\eqs{nS}{aS}, we can extract the values for $n_{\rm s}, \as$ and
$r$, thereby testing our model against the observational data of
\eqs{nswmap}{obs3}.

\begin{figure}[!t]
\includegraphics[width=60mm,angle=-90]{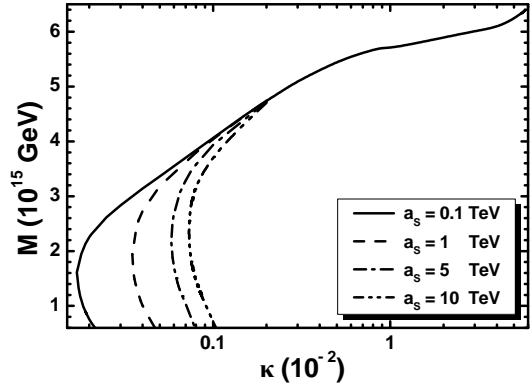}
\vspace*{-0.5cm} \caption{\label{kM}\sl
$M$ versus $\kappa$ for various $\aS$ values.}
\end{figure}

Our results are presented in Figs.~\ref{fig1}, \ref{arns} and
\ref{kM} taking $\aS=0.1~\TeV$ (solid lines), $\aS=1~{\rm TeV}$
(dashed lines), $\aS=5~{\rm TeV}$ (dot-dashed lines), and
$\aS=10~{\rm TeV}$ (double dot-dashed lines). In Figs.~\ref{fig1}
and \ref{arns} the observationally compatible region of
Eq.~(\ref{nswmap}) is also indicated by the horizontal (in
\Fref{fig1}) or vertical (in \Fref{arns}) lines. For the sake of
clarity, we do not show solutions with $M>2\cdot10^{16}~\GeV$ --
cf. \cref{sstad2} -- which are totally excluded by \Eref{nswmap}.

From \Fref{fig1}, where we depict $n_{\rm s}$ versus $\kappa$ {\sf
\small (a)} and $M$ {\sf\small (b)}, we note that, for
$\kappa\gtrsim0.002$ and $M\gtrsim4.7\cdot10^{15}~\GeV$, $V_{\rm
HIc}$ and progressively -- for $\kappa\gtrsim0.04$ and
$M\gtrsim6.1\cdot10^{15}~\GeV$, -- $V_{\rm HIS}$ dominates $\Vhi$
in \Eref{Vol}, and drives $n_{\rm s}$ to values close to or larger
than $0.98$, independently of the selected $\aS$ values. On the
other hand, for $\kappa\lesssim0.002$, $V_{\rm HIT}$ starts
becoming comparable to $V_{\rm HIc}$ and succeeds in reconciling
$\ns$ with \Eref{nswmap} for well defined $\kappa$ (and $M$)
values that are related to the chosen $\aS$. Actually, for the
allowed $\ns$, we find that $V_{\rm HIc}/V_{\rm HIT}\simeq13$,
whereas $V_{\rm HIS}$ turns out to be totally negligible. Fixing
$\ns$ to its central value in \Eref{nswmap}, we display in
\Tref{tab1} the values for $(\kp,M)$ corresponding to the $\aS$
values employed in Figs.~\ref{fig1}-\ref{kM}.

From our numerical computations we observe that, in the regime
with acceptable $\ns$ values, the $\sg_*$ required by
\eqs{Nhi}{Prob} becomes comparable to $\sg_{\rm c}$, and $f_{\rm
rc}(x)$ in \Eref{frc} can be approximated as \cite{hinova}
\bea \nonumber f_{\rm rc}(x)&\simeq&3 - {x^{-2}\over6} -
{x^{-4}\over30} - {x^{-6}\over84} - {x^{-8}\over180} -
{x^{-10}\over330} \\ &-& {x^{-12}\over546} - {x^{-14}\over840} -
{x^{-16}\over1224} - {x^{-18}\over1710} -
{x^{-20}\over2310}\cdot~~\label{frc2}\eea
Moreover, in the vicinity of $\sg_*$, $\Vhi$ develops a local
maximum at $\sgmax$ allowing for FHI of hilltop type \cite{lofti}
to take place. As a consequence, $\Vhi'$, and therefore $\epsilon$
in \Eref{slow} and $r$ in \Eref{aS} -- see \sFref{arns}{b} --,
decrease sharply (enhancing $\Nhi$), whereas $|\Vhi''|$ (or
$|\eta|$) increases adequately, thereby lowering $\ns$ within the
range of \Eref{nswmap}. In particular, for constant $\kp$, the
lower the value for $n_{\rm s}$ we wish to attain, the closer we
must set $\sgex$ to $\sgmax$. To quantify the amount of these
tunings, we define the quantities
\beq
\Dex={\sgex-\sgc\over\sgc}~~\mbox{and}~~\Dmax={\sgmax-\sgex\over\sgmax}\label{dms}\eeq
and list their resulting values in \Tref{tab1}.  From there, we
conclude that the required tuning is at a few percent level, since
$\Dex,\Dmax\leq10\%$. Values of $\aS$ well below $1~\TeV$ are less
desirable from this point of view. For comparison, we mention that
for $\kp\geq0.002$, we get $\Dex\geq30\%$, i.e., $\Dex$ increases
with $\kp$ whereas the maximum disappears. From \Tref{tab1}, we
note that $\kp$ and $M$ decrease with $\Dex$ and $\Dmax$, too.

\begin{table}[!t]
\caption{\sl Model parameters and predictions for
$\ns\simeq0.96$.}
\begin{tabular}{c@{\hspace{0.3cm}}c@{\hspace{0.3cm}}c@{\hspace{0.3cm}}c@
{\hspace{0.3cm}}c@{\hspace{0.3cm}}c@{\hspace{0.3cm}}c} \toprule
{$\aS$} &{$\kp $}&{$M$}&{$\Dex$}&{$\Dmax$}&$-\as$&$r$\\
{$(\TeV)$} &{$
(10^{-4})$}&{$(10^{15}~\GeV)$}&{$(\%)$}&{$(\%)$}&$(10^{-4})$&$
(10^{-13})$\\\colrule
{$0.1$} &{$2.05$}&{$0.7$}&{$0.6$}&$0.016$&$1.5$&$0.09$\\
{$1$} &{$3.9$}&{$1.1$}&{$2$}&{$1.2$}&$1.9$&$1.9$\\
{$5$} &{$6.3$}&{$1.4$}&{$4.3$}&{$2.8$}&$2.4$&$15$\\
{$10$} &{$7.7$}&{$1.6$}&{$6.3$}&{$3.8$}&$2.5$&$38$\\\botrule
\end{tabular}
\label{tab1}
\end{table}

In \sFref{arns}{a} and \sFref{arns}{b} respectively we display the
predictions of our model for $|\as|\equiv |d\ns / d\ln k|$ and
$r$. Corresponding to the $\ns$ values within \Eref{nswmap},
$|\as|$ turns out to be of order $10^{-4}$. On the contrast, $r$
is extremely tiny, of order $10^{-14} - 10^{-12}$, and therefore
far outside the reach of PLANCK and other contemporary
experiments. For the preferred $\ns$ values, we observe that $r$
and $|\as|$ increase with $\aS$ whereas for constant $\aS$, $\as$,
and $r$ increase with $\ns$. For the $\aS$ values used in
\Fref{arns} and with $\ns=0.96$, our predictions are summarized in
\Tref{tab1}.

The dependence of $M$ on $\kp$ within our model is shown in
\Fref{kM}. We remark that $M$ mostly decreases with $\kp$. For low
enough $\kp$ values, there is region where we get two $M$ values
consistent with \eqs{Nhi}{Prob}. Comparing \Fref{kM} with
\sFref{fig1}{b}, we can easily conclude that the latter solution
is consistent with \Eref{nswmap}. The $M$ values displayed in this
figure are fully compatible with the upper bound arising from
\Eref{mucs}. Although these $M$ values lie somewhat below $\Mgut$,
the unification of gauge coupling constants within MSSM remains
intact since the gauge boson associated with the spontaneous
$U(1)_{B-L}$ breaking is neutral under $G_{\rm MSSM}$, and so it
does not contribute to the relevant \emph{renormalization group}
(RG) running.

\begin{figure}[!b]
\includegraphics[width=60mm,angle=-90]{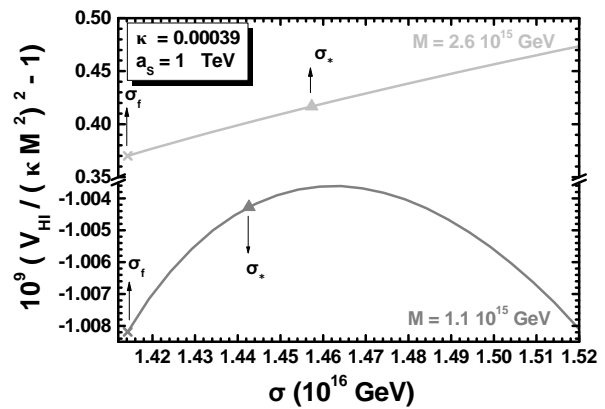}
\vspace*{-0.5cm} \caption{\label{Vhi}\sl The variation of $\Vhi$
as a function of $\sg$ for $\aS=1~\TeV$, $\kappa=3.9\cdot 10^{-4}$
and $M=1.1\cdot10^{15}~\GeV$ ($\ns=0.96$, gray line) or
$M=2.6\cdot10^{15}~\GeV$ ($\ns=0.994$, light gray line).}
\end{figure}

In order to highlight the differences of the various possible
solutions obtained at low $\kp$ values, we present in
Fig.~\ref{Vhi} the variation of $\Vhi$ as a function of $\sg$ for
the same $\kappa$ and $\aS$ and two different $M$ values
compatible with \eqs{Nhi}{Prob}. Namely, we take $\aS=1~\TeV$,
$\kappa=3.9\cdot 10^{-4}$ and $M=1.1\cdot10^{15}~\GeV$
[$M=2.6\cdot10^{15}~\GeV$] yielding $\ns=0.96$ [$\ns=0.994$] with
$\Dex=2\%$ [$\Dex=3\%$] -- gray [light gray] line. The
corresponding  $\sigma_*$ and $\sigma_{\rm f}$ values are also
shown. As we anticipated above, in the first case, $\Vhi$ develops
a maximum at $\sigma_{\rm max}\simeq1.46M$ decreasing thereby
$\ns$ at an acceptable level -- we get $\Dmax=1.2\%$ as shown in
\Tref{tab1}. Needless to say that, in both cases, $\Vhi$ turns out
to be bounded from below for large $\sg$ values and, therefore, no
complications arise in the realization of the inflationary
dynamics.

\begin{figure*}[!t]
\centering
\includegraphics[width=60mm,angle=-90]{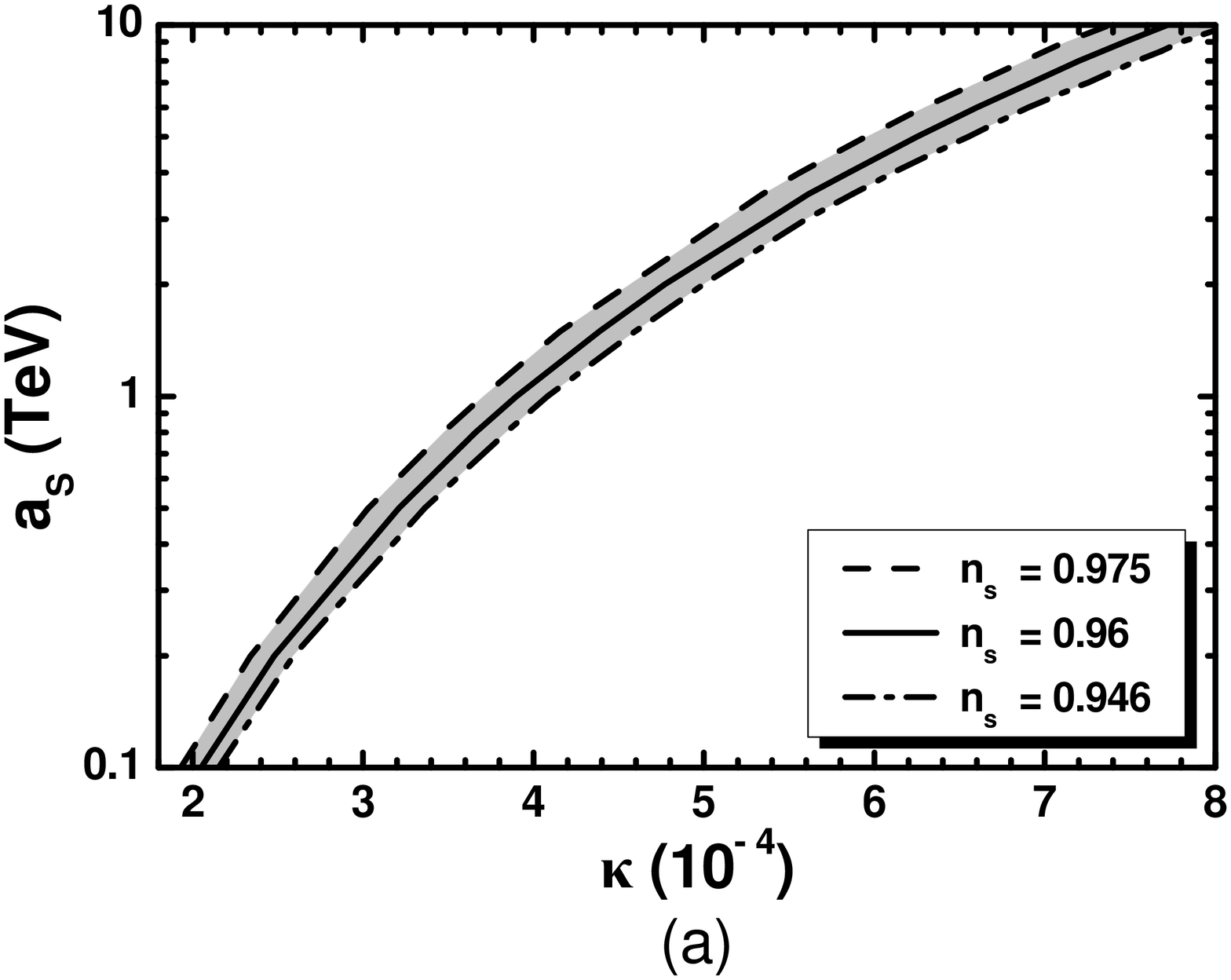}
\includegraphics[width=60mm,angle=-90]{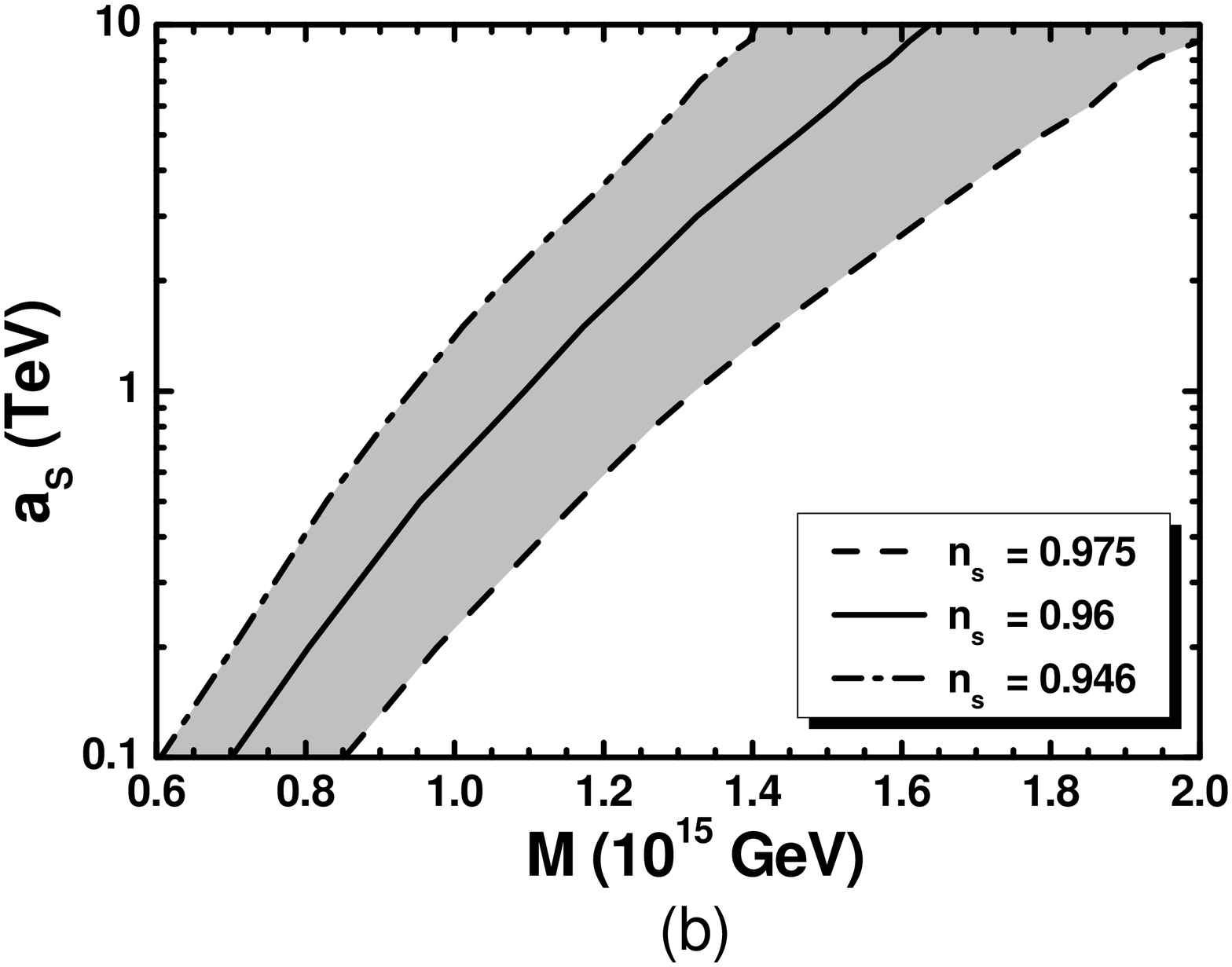}
\caption{\label{fig2}\sl Allowed (shaded) regions as determined by
Eqs.~(\ref{Nhi}), (\ref{Prob}), (\ref{nswmap}) and (\ref{mucs}) in
(a) $\kappa-\aS$ plane and (b) $M-\aS$ plane. The $n_s$ values for
the various lines are also shown. }
\end{figure*}

As inferred from \Fref{fig1}, for any $\kp\lesssim10^{-4}$ we can
conveniently adjust $\aS$, so that \Eref{nswmap} is fulfilled.
Working in this direction, we delineate the (lightly gray) region
in the $\kappa-\aS$ [$M-\aS$] plane allowed by all the imposed
constraints -- see \sFref{fig2}{a} [\sFref{fig2}{b}]. We also
display by solid lines the allowed contours for $\ns=0.96$. We do
not consider $\aS$ values lower than $0.1~\TeV$, since they would
be less natural from the point of view of both SUSY breaking and
the $\Dex$'s and $\Dmax$'s encountered -- see \Tref{tab1}. The
boundaries of the allowed areas in \Fref{fig2} are determined by
the dashed [dot-dashed] lines corresponding to the lower [upper]
bound on $n_{\rm s}$ in Eq.~(\ref{nswmap}). In these regions we
obtain $mu_{\rm cs}=(0.98-12.4)\cdot10^{-7}\mP^2$ which are
compatible with Eq.~(\ref{mucs}). On the other hand, these regions
are not consistent with the most stringent (although controversial
\cite{sanidas}) constraint $\mu_{\rm cs}\lesssim10^{-7}\mP^2$
\cite{pta} imposed by the limit on the stochastic gravitational
wave background from the European Pulsar Timing Array. These
latter results depend on assumptions regarding string loop
formation and the gravitational waves emission. The bounds on $M$
from $\mu_{\rm cs}$, are totally avoided if we implement FHI
within $G_{\rm LR}$ \cite{dvali} or flipped $SU(5)$
\cite{flipped}, with ${\sf N}=2$ or ${\sf N}=10$ respectively in
\Eref{Vcor}, which do not lead to the production of any cosmic
defect -- for a more complete discussion involving flipped $SU(5)$
and the corresponding $M$ values, see second paper in
\cite{sstad2}.

Summarizing our findings from \Fref{fig2}, for $n_{\rm s}$
considered by \Eref{nswmap} and $0.1\lesssim{\aS/\TeV}\lesssim10$,
we obtain:
\beqs\bea &&1.9\lesssim{\kp/10^{-4}}\lesssim8.1,~~0.6\lesssim
{M/10^{15}~{\rm GeV}}\lesssim2, ~~~~\label{res1}  \\
\label{res2} && ~1.1\lesssim {|\as|/10^{-3}}\lesssim2.8,~~
0.05\lesssim {r/10^{-13}}\lesssim76.~~~~~~~~~\eea\eeqs
The $M$ values are consistent with \Eref{mucs} according to which
$M\lesssim(5-5.45)\cdot10^{15}~\GeV$. The maximal values for
$|\as|$ and $r$ are respectively encountered in the upper left and
right corners of the allowed region in \sFref{fig2}{b}. In the
lower left [right] corner of that area, we obtain the lowest
possible $r$ [$|\as|$]. Also, $\Dex$ ranges between $0.6\%$ and
$7.3\%$ whereas $\Dmax$ varies between $0.001\%$ and $7.9\%$.

\section{Non-Thermal Leptogenesis}\label{pfhi}

\setcounter{paragraph}{0}

\paragraph{\small\sf\scshape  Inflaton Decay.} As \FHI ends,
$\sg$ crosses $\sgc$, thereby destabilizing the $\Phi-\bar\Phi$
system which leads to a stage of tachyonic preheating as described
in \cref{buch}. Soon afterwards, the \emph{inflaton system} (IS)
settles into a phase of damped oscillations about the SUSY vacuum,
eventually decaying and reheating the universe. Note that the IS
consists of the two complex scalar fields $S$ and $(\delta\bar
\Phi+\delta \Phi)/\sqrt{2}$, where $\delta\bar \Phi=\bar \Phi-M$
and $\delta\Phi=\Phi-M$.  To ensure the decay of the IS and
implement the see-saw mechanism for the generation of the light
neutrino masses, we allow for the following superpotential terms:
\beq W_{\rm RHN}=\ld_i\bar\Phi \nu^c_i \nu^c_i +h_{Nij} \sni L_j
\hu, \label{Wrhn}\eeq
where $\bar\Phi$ [$\sni$] have $B-L$ charge of $-2[1]$ and R
charge $0$ $[\alpha/2]$. $L_i$ denotes the $i$-th generation
$SU(2)_{\rm L}$ doublet left-handed lepton superfields, and $\hu$
is the $SU(2)_{\rm L}$ doublet Higgs superfield which couples to
the up quark superfields.

At the SUSY vacuum, \Eref{vevs}, $\Phi$ and $\bar\Phi$ acquire
their v.e.vs, thereby providing masses to the IS and $\nu_i^c$'s,
\beq \label{masses} \mbox{\small\sf(a)}\>\>\>
\msn=\sqrt{2}{\kp\Mpq}\>\>\>\mbox{and}\>\>\>\mbox{\small\sf
(b)}\>\>\>\mrh[i]=2{\ld_{i}\Mpq}.\eeq
The predominant decay channels of $S$ and $(\delta\bar \Phi+\delta
\Phi)/\sqrt{2}$ are to (kinematically allowed) bosonic and
fermionic $\sni$'s respectively via tree-level couplings derived
from \eqs{Whi}{Wrhn} -- see e.g. \cref{lectures} -- with almost
the same decay width \cite{gmb}
\beq \Gm[{\rm I}\to \nu^c_i]={1\over64\pi}\lambda_i
^2\,\msn\sqrt[3]{1-{4\mrh[i]^2/\msn^2}}. \label{gammas}\eeq
We assume here that the $\mu$ problem of MSSM is resolved as
suggested in \cref{flipped,rsym}, rather than by invoking the
mechanism of \cref{dvali} which would open new and efficient decay
channels for $S$. The SUGRA-induced \cite{Idecay} decay channels
are negligible in our set-up, with the $M$ and $\msn$ values in
\Eref{res1}. The resulting reheat temperature is given by
\cite{quin}
\beq T_{\rm rh}\approx \left(72/5\pi^2g_{\rm *}\right)^{1/4}
\sqrt{\mbox{$\sum_i$}\Gm[{\rm I}\to \nu^c_i]\ m_{\rm
P}},\label{Trh}\eeq
where $g_{*}=228.75$ counts the MSSM effective number of
relativistic degrees of freedom at temperature $T_{\rm rh}$.

For $\Trh<\mrh[i]$, the out-of-equilibrium decay of  $\nu^c_{i}$
generates a lepton-number asymmetry (per $\nu^c_{i}$ decay),
$\ve_i$. The resulting lepton-number asymmetry is partially
converted through sphaleron effects into a yield of the observed
BAU:
\beq Y_B=-0.35\cdot2\cdot{5\over4}{\Trh\over\msn}\mbox{$\sum_i$}
\br_i\ve_i , \>\>\>\mbox{with}\>\>\>\br_i={\Gm[{\rm I}\to
\nu^c_i]\over\sum_i\Gm[{\rm I}\to \nu^c_i]} \label{Yb}\eeq
being the branching ratio of IS to $\sni$. The quantity $\ve_i$
can be expressed in terms of the Dirac masses of $\nu_i$,
$\mD[i]$, arising from the second term of \Eref{Wrhn}.

The required $\Trh$ in \Eref{Yb} must be compatible with
constraints on the gravitino ($\Gr$) abundance, $Y_{3/2}$, at the
onset of \emph{nucleosynthesis} (BBN), which is estimated to be
\cite{kohri}:
\beq\label{Ygr} Y_{3/2}\simeq 1.9\cdot10^{-22}\ \Trh/\GeV ,\eeq
where we take into account only thermal production of $\Gr$, and
assume that $\Gr$ is much heavier than the MSSM gauginos -- the
case of $\Gr$ CDM was recently analyzed in \cref{buch}.

\paragraph{\small\sf\scshape  Post-Inflationary Requirements.} The
success of our post-inflationary scenario can be judged, if, in
addition to the constraints of \Sref{fhi}, it is consistent with
the following requirements:

$\bullet$ The bounds on $\mrh[i]$:
\beq\label{kin} \mrh[i]\lesssim 7.1 M,\>\>\mrh[1]\gtrsim
10\Trh\>\>\mbox{and}\>\>\msn\geq2\mrh[i],\eeq
for some  $\nu^c_{i}$'s. The first bound comes from the needed
perturbativity of $\ld_i$'s in \Eref{Wrhn}, i.e.
$\ld_{i}\leq\sqrt{4\pi}$. The second inequality is applied to
avoid any erasure of the produced $Y_L$ due to $\nu^c_1$ mediated
inverse decays and $\Delta L=1$ scatterings \cite{lsenoguz}.
Finally, the last bound above ensures a kinematically allowed decay
of the IS for some $\sni$'s.

$\bullet$ Constraints from Neutrino Physics. We take as inputs the
best-fit values \cite{valle} -- see also \cref{lisi} -- on the
neutrino mass-squared differences, $\Delta
m^2_{21}=7.62\cdot10^{-3}~{\rm eV}^2$ and $\Delta m^2_{31}=\lf
2.55\left[-2.43\right]\rg\cdot~10^{-3}~{\rm eV}^2$, on the mixing
angles, $\sin^2\theta_{12}=0.32$,
$\sin^2\theta_{13}=0.0246\left[0.025\right]$, and
$\sin^2\theta_{23}=0.613\left[0.6\right]$ and the CP-violating
Dirac phase $\delta=0.8\pi\left[-0.03\pi\right]$ for \emph{normal
[inverted] ordered} (NO [IO]) \emph{neutrino masses}, $\mn[i]$'s.
The sum of $\mn[i]$'s is bounded from above by the data
\cite{wmap, plcp}, $\sum_i \mn[i]\leq0.28~{\eV}$ at 95\% c.l.

$\bullet$ The observational results on $Y_B$ \cite{wmap, plcp}
\beq Y_B\simeq\lf8.55\pm0.217\rg\cdot10^{-11}~~\mbox{at 95\%
c.l.}\label{BAUwmap}\eeq

$\bullet$ The bounds on $Y_{3/2}$ imposed \cite{kohri} by
successful BBN:
\beq  \label{Ygw} Y_{3/2}\lesssim\left\{\bem
%
10^{-14}\hfill \cr
4.3\cdot10^{-14}\hfill \cr
10^{-13}\hfill \cr\eem
\right.\>\>\>\mbox{for}\>\>\>m_{3/2}\simeq\left\{\bem
0.69~{\rm TeV}\hfill \cr
8~{\rm TeV}\hfill \cr
10.6~{\rm TeV.}\hfill \cr\eem
\right.\eeq
Here we consider the conservative case where $\Gr$ decays with a
tiny hadronic branching ratio.

\paragraph{\small\sf\scshape  Results.}


\renewcommand{\arraystretch}{1.4}
\begin{table}[!t]
\caption{\sl Parameters yielding the correct BAU for
$\kp=0.00039$, $\aS=1~\TeV$ and various neutrino mass schemes.}
\begin{tabular}{c||c|c||c|c|c||c|c}\toprule
Parameters &  \multicolumn{7}{c}{Cases}\\\cline{2-8}
&A&B& C & D& E & F&G\\ \cline{2-8} &\multicolumn{2}{c||}{Normal} &
\multicolumn{3}{|c||}{Degenerate}&  \multicolumn{2}{|c}{Inverted}
\\& \multicolumn{2}{c||}{Hierarchy}&\multicolumn{3}{|c||}{Masses}& \multicolumn{2}{|c}{Hierarchy}\\
\colrule
\multicolumn{8}{c}{Low Scale Parameters}\\\colrule
$\mn[1]/0.1~\eV$&$0.01$&$0.1$&$0.5$ & $0.7$& $0.7$ & $0.5$&$0.49$\\
$\mn[2]/0.1~\eV$&$0.09$&$0.13$&$0.51$ & $1.0$& $0.705$ & $0.51$&$0.5$\\
$\mn[3]/0.1~\eV$&$0.5$&$0.51$&$0.71$ & $1.12$&$0.5$ &
$0.1$&$0.05$\\\colrule
$\sum_i\mn[i]/0.1~\eV$&$0.6$&$0.74$&$1.7$ & $2.3$&$1.9$ &
$1.1$&$1$\\ \colrule
$\varphi_1$&$0$&$\pi/3$&$0$ & $\pi/2$&$0$ & $-\pi/6$&$0$\\
$\varphi_2$&$0$&$0$ &$\pi/3$& $0$&$-\pi/2$ &
$0$&$-\pi/3$\\\colrule
\multicolumn{8}{c}{Leptogenesis-Scale Parameters}\\\colrule
$\mD[1]/0.1~\GeV$&$1.67$&$4.1$&$3.7$ & $7$&$7$ & $5$&$60$\\
$\mD[2]/\GeV$&$4$&$0.5$&$1.1$ & $1.55$&$1.03$ & $0.93$&$4$\\
$\mD[3]/\GeV$&$120$&$120$&$5$ & $2$&$2$ & $4$&$1.32$\\\colrule
$\mrh[1]/10^{9}~\GeV$&$2.5$&$2.4$&$3.3$ & $6.5$&$4.6$ & $1$&$48$\\
$\mrh[2]/10^{10}~\GeV$& $47$&$1.6$&$1.7$&$2.7$ &$1.6$ & $2.8$&$59$\\
$\mrh[3]/10^{12}~\GeV$&$3720$&$580$&$0.34$ & $0.035$&$0.046$ &
$0.7$&$10$\\\colrule
\multicolumn{8}{c}{Decay channels of the Inflaton System,
I}\\\colrule
I $\to$&$\wrhn[1]$&$\wrhn[1,2]$& $\wrhn[1,2]$& $\wrhn[1,2,3]$&
$\wrhn[1,2,3]$ & $\wrhn[1,2]$&$\wrhn[1]$\\ \colrule
\multicolumn{8}{c}{Resulting $B$-Yield }\\\colrule
$10^{11}Y^0_B$&$8.9$&$8.25$& $8$& $6$&$6.9$ & $8.3$&$11.1$\\
$10^{11}Y_B$&$8.5$&$8.6$& $8.6$& $8.6$&$8.5$ &
$8.5$&$8.6$\\\colrule
\multicolumn{8}{c}{Resulting $\Trh$ and $\Gr$-Yield }\\\colrule
$\Trh/10^{8}~\GeV$&$0.7$&$2$& $1.9$& $4.1$&$5.5$ & $3$&$5$\\
$10^{14}Y_{3/2}$&$1.3$&$3.8$& $3.6$& $9.5$&$10$ &
$6$&$10$\\\botrule
\end{tabular}
 \label{tab4}
\end{table}

The inflationary requirements of \Sref{fhi} restrict $\kp$ and $M$
in the very narrow range presented in \Eref{res1}. As a
consequence, the mass $\msn$ of IS given by \Eref{masses}, is
confined to the range $(2-17.8)\cdot10^{11}~\GeV$, and its
variation is not expected to decisively influence our results on
$\Yb$. For this reason, throughout our analysis here we use the
central value $\msn\simeq6\cdot10^{11}~\GeV$, corresponding to the
second row of \Tref{tab1}.

On the other hand, $\Trh$ (and $Y_B$) also depend on the masses
$\mrh[i]$ of $\sni$ into which the IS decays. Following the
bottom-up approach -- see Sec.~IVB of \cref{nMCI} --, we find the
$\mrh[i]$'s by using as inputs the $\mD[i]$'s, a reference mass of
the $\nu_i$'s -- $\mn[1]$ for NO $\mn[i]$'s, or $\mn[3]$ for IO
$\mn[i]$'s --, the two Majorana phases $\varphi_1$ and $\varphi_2$
of the MNS matrix, and the best-fit values mentioned above for the
low energy parameters of neutrino physics. In our numerical code,
we also estimate, following \cref{running}, the RG evolved values
of the latter parameters at the scale of nTL, $\Lambda_L=\msn$, by
considering the MSSM with $\tan\beta\simeq50$ as an effective
theory between $\Lambda_L$ and the SUSY-breaking scale, $M_{\rm
SUSY}=1.5~\TeV$. We evaluate the $\mrh[i]$'s at $\Lambda_L$, and
we neglect any possible running of the $\mD[i]$'s and $\mrh[i]$'s.
Therefore, we present their values at $\Lambda_L$.

Our results are displayed in \Tref{tab4} taking some
representative values of the parameters which yield the correct
$\Yb$, as dictated by \Eref{BAUwmap}. We consider NO (cases A and
B), degenerate (cases C, D and E) and IO (cases F and G)
$\mn[i]$'s. In all cases the current limit (see point 2 above) on
the sum of $\mn[i]$'s is safely met -- the case D approaches it.
The gauge group adopted here, $G_{B-L}$, does not predict any
relation between the Yukawa couplings constants $h_N$ entering the
second term of \Eref{Wrhn} and the other Yukawa couplings in the
MSSM. As a consequence, the $\mD[i]$'s are free parameters.
However, for the sake of comparison, for case A, we take
$\mD[3]=m_t(\Lambda_L)$, and in case B, we also set
$\mD[2]=m_c(\Lambda_L)$, where $m_t$ and $m_c$ denote the masses
of the top and charm quark respectively. We observe that in all
cases $\mD[1]\gtrsim0.1~\GeV$. This is done, in order to fulfill
the second inequality in \Eref{kin}, given that $\mD[1]$ heavily
influences $\mrh[1]$. Note that such an adjustment requires
theoretical motivation, if the gauge group is $G_{\rm LR}$ or
flipped $SU(5)$ -- cf. \cref{lsenoguz}.

From \Tref{tab4} we observe that with NO or IO $\mn[i]$'s, the
resulting $\mrh[i]$'s are also hierarchical. With degenerate
$\mn[i]$'s, the resulting $M_{i\nu}$'s are closer to one another.
Therefore, in the latter case more IS-decay channels are
available, whereas for cases A and G only a single decay channel
is open. In all other cases, the dominant contributions to $\Yb$
arise from $\ve_2$. In \Tref{tab4} we also display, for
comparison, the $B$-yield with ($Y_B$) or without ($Y^0_B$) taking
into account the RG effects. We observe that the two results are
mostly close to each other with some discrepancies appearing for
degenerate and IO $\mn[i]$'s. Shown also are values for $\Trh$,
the majority of which are close to $5\cdot10^8~\GeV$, and the
corresponding $Y_{3/2}$'s, which are consistent with \Eref{Ygw}
mostly for $m_{3/2}\gtrsim8~\TeV$. These large values can be
comfortably tolerated with the $\aS$'s appearing in \Fref{fig2}
for $A\sim1$ -- see the definition of $\aS$ below \Eref{Vtad}.
From the perspective of $\Gr$ constraint, case A turns out to be
the most promising.

\section{Conclusions}\label{con}

Inspired by the recently released WMAP and PLANCK results for the
inflationary observables, we have reviewed and updated the
predictions arising from a minimal model of SUSY (F-term) hybrid
inflation, also referred to as FHI. In this set-up
\cite{susyhybrid}, FHI is based on a unique renormalizable
superpotential, employs a canonical \Ka, and is associated with a
superheavy $B-L$ phase transition. As shown in \cref{sstad2}, and
verified by us here, to achieve $\ns$ values lower than $0.98$,
one should include in the inflationary potential the soft SUSY
breaking tadpole term, with the SUSY breaking mass parameter
values in the range $(0.1 - 10)~\TeV$. Fixing $n_{\rm s}$ to its
central value, the dimensionless coupling constant, the $B-L$
symmetry breaking scale, and the inflationary parameters $\as$ and
$r$ are respectively given by $\kp=(2-7.7)\cdot10^{-4}$,
$M=(0.7-1.6)\cdot10^{15}~\GeV$, $|\as| \simeq
(1.5-2.5)\cdot10^{-4}$ and $r\simeq(0.1-37)\cdot10^{-13}$. The
$B-L$ cosmic strings, formed at the end of FHI, have tension
ranging from  $1.3$ to $8.3\cdot10^{-7}\mP^2$ and may be
accessible to future observations. We have also briefly discussed
the reheat temperature, gravitino constraints and non-thermal
leptogenesis taking into account updated values for the neutrino
oscillation parameters.

\acknowledgments  Q.S. acknowledges support by the DOE grant No.
DE-FG02-12ER41808. We would like to thank W. Buchm\"uller,
M.~Hindmarsh, A.~Mazumdar and K. Schmitz for useful discussions.


\def\ijmp#1#2#3{{\sl Int. Jour. Mod. Phys.}
{\bf #1},~#3~(#2)}
\def\plb#1#2#3{{\sl Phys. Lett. B }{\bf #1}, #3 (#2)}
\def\prl#1#2#3{{\sl Phys. Rev. Lett.}
{\bf #1},~#3~(#2)}
\def\rmp#1#2#3{{Rev. Mod. Phys.}
{\bf #1},~#3~(#2)}
\def\prep#1#2#3{{\sl Phys. Rep. }{\bf #1}, #3 (#2)}
\def\prd#1#2#3{{\sl Phys. Rev. D }{\bf #1}, #3 (#2)}
\def\npb#1#2#3{{\sl Nucl. Phys. }{\bf B#1}, #3 (#2)}
\def\npps#1#2#3{{Nucl. Phys. B (Proc. Sup.)}
{\bf #1},~#3~(#2)}
\def\mpl#1#2#3{{Mod. Phys. Lett.}
{\bf #1},~#3~(#2)}
\def\jetp#1#2#3{{JETP Lett. }{\bf #1}, #3 (#2)}
\def\app#1#2#3{{Acta Phys. Polon.}
{\bf #1},~#3~(#2)}
\def\ptp#1#2#3{{Prog. Theor. Phys.}
{\bf #1},~#3~(#2)}
\def\n#1#2#3{{Nature }{\bf #1},~#3~(#2)}
\def\apj#1#2#3{{Astrophys. J.}
{\bf #1},~#3~(#2)}
\def\mnras#1#2#3{{MNRAS }{\bf #1},~#3~(#2)}
\def\grg#1#2#3{{Gen. Rel. Grav.}
{\bf #1},~#3~(#2)}
\def\s#1#2#3{{Science }{\bf #1},~#3~(#2)}
\def\ibid#1#2#3{{\it ibid. }{\bf #1},~#3~(#2)}
\def\cpc#1#2#3{{Comput. Phys. Commun.}
{\bf #1},~#3~(#2)}
\def\astp#1#2#3{{Astropart. Phys.}
{\bf #1},~#3~(#2)}
\def\epjc#1#2#3{{Eur. Phys. J. C}
{\bf #1},~#3~(#2)}
\def\jhep#1#2#3{{\sl J. High Energy Phys.}
{\bf #1}, #3 (#2)}
\newcommand\jcap[3]{{\sl J.\ Cosmol.\ Astropart.\ Phys.\ }{\bf #1}, #3 (#2)}
\newcommand\njp[3]{{\sl New.\ J.\ Phys.\ }{\bf #1}, #3 (#2)}


\begin{thebibliography}{99}
\section*{\refname} \ignorespaces

\bibitem{susyhybrid} G.R. Dvali, Q. Shafi and R.K. Schaefer, \prl{73}{1994}{1886} [{\ftn\tt
hep-ph/9406319}].


\bibitem{hybrid} E.J. Copeland {\it et al.}, {\sl Phys. Rev. D }{\bf 49}, {6410} ({1994})
[{\ftn\tt astro -ph/9401011}].

\bibitem{bl} V.N.~\c{S}eno\u{g}uz and Q.~Shafi, {\tt\ftn hep-ph/0512170}.


\bibitem{dvali} G.R. Dvali, G. Lazarides and Q. Shafi, {\sl Phys.
Lett. B} {\bf 424}, {259} ({1998}) [\hepph{9710314}].


\bibitem{flipped} B.~Kyae and Q.~Shafi, \plb{635}{2006}{247} [\hepph{0510105}].

\bibitem{sstad2} M.U.~Rehman, Q.~Shafi and J.R.~Wickman, {\sl Phys.\ Lett.\ B} {\bf
683}, 191 (2010) [\arxiv{0908.3896}]; M.~U.~Rehman, Q.~Shafi and
J.~R.~Wickman, {\sl Phys.\ Lett.\ B }{\bf 688}, 75 (2010)
[{\tt\ftn arXiv:0912.4737}]; M.~Civiletti, M.~U.~Rehman, E.~Sabo,
Q.~Shafi and J.~Wickman, {\tt\ftn arXiv:1303.3602}.



\bibitem{senoguz} A.D. Linde and A. Riotto \prd{56}{1997}{1841} [{\ftn\tt hep-ph/9703209}];
V.N. \c{S}eno\u{g}uz and Q. Shafi, \plb{567}{2003}{79} [{\ftn\tt
hep-ph/0305089}].



\bibitem{sstad1} V.N. \c{S}eno\u{g}uz and Q. Shafi, {\sl Phys.\ Rev.\ D}~{\bf
71}, 043514 (2005) [{\ftn\tt hep-ph/0412102}].



\bibitem{gpp} M. Bastero-Gil, S.F. King, and Q. Shafi, \plb{651}{2007}{345}
[{\ftn\tt hep-ph/0604198}]; B. Garbrecht \etal,
\jhep{12}{2006}{038} [{\ftn\tt hep-ph/0605264}]; M.U. Rehman,
V.N.~\c{S}eno\u{g}uz, and Q.~Shafi, {\sl Phys. Rev. D }{\bf 75},
043522 (2007) [{\ftn\tt hep-ph/0612023}]; C. Pallis,
\jcap{04}{2009}{024} [\arxiv{0902.0334}].

\bibitem{rlarge} M.U.~Rehman, Q.~Shafi and J.R.~Wickman, {\sl Phys.\ Rev.\ D} {\bf
83}, 067304 (2011) [\arxiv{1012.0309}].

\bibitem{hinova} R. Armillis and C. Pallis, {\sl ``Recent Advances in Cosmology''},
edited by A. Travena and B. Soren (Nova Science Publishers Inc.,
New York, 2013)  [\arxiv{1211.4011}].

\bibitem{mhi} G. Lazarides and C. Pallis, \plb{651}{2007}{216}
[{\tt\ftn hep- ph/0702260}].


\bibitem{wmap} G. Hinshaw \etal\ [WMAP Collaboration], \arxiv{1212.5226}.

\bibitem{plin} P.A.R.~Ade {\it et al.}  [Planck Collaboration],
\arxiv{1303.5082}.

\bibitem{plcp} P.A.R.~Ade {\it et al.}  [Planck Collaboration],
\arxiv{1303.5076}.





\bibitem{lept}  G. Lazarides and Q. Shafi, \plb{258}{1991}{305};\\
K. Kumekawa, T. Moroi and T. Yanagida, {\sl Prog. Theor. Phys.
}{\bf 92}, 437 (1994) [\hepph{9405337}]; G. Lazarides, R.K.
Schaefer and Q. Shafi, \prd{56}{1997}{1324} [{\ftn\tt
hep-ph/9608256}]; V.N.~\c{S}eno\u{g}uz and Q.~Shafi, {\sl Phys.\
Rev.\ D }{\bf 71}, 043514 (2005)  [{\tt\ftn hep-ph/0412102}].


\bibitem{gravitino} M.Yu. Khlopov and A.D. Linde,
{\sl Phys. Lett. B }{\bf 138}, 265 (1984);  J. Ellis, J.E. Kim,
and D.V. Nanopoulos, \ibid{145}{1984}{181}.

\bibitem{kohri} M.Kawasaki, K.Kohri and T.Moroi, {\sl Phys. Lett. B}~{\bf 625}, {7}
({2005}) [\astroph{0402490}]; M.~Kawasaki, K.~Kohri and T.~Moroi,
\prd{71}{2005}{083502} [\astroph{0408426}]; R.H.~Cyburt \etal,
\prd{67}{2003}{103521} [{\ftn\tt astro- ph/0211258}]; J.R.~Ellis,
K.A.~Olive and E.~Vangioni, \plb{619}{2005}{30}
[\astroph{0503023}].

\bibitem{valle}  D.V.~Forero, M.~Tortola and J.W.F.~Valle,
  {\sl Phys.\ Rev.\ D }{\bf 86}, 073012 (2012) [\arxiv{1205.4018}].

\bibitem{lisi} G.L.~Fogli \etal, {\sl Phys.\ Rev.\ D }{\bf 86}, 013012 (2012)
[{\ftn\tt arXiv: 1205.5254}].



\bibitem{plcs} P.A.R. Ade {\it et al.}  [Planck Collaboration],
\arxiv{1303.5085}.

\bibitem{strings}  M.~Hindmarsh, {\sl Prog.\ Theor.\ Phys.\ Suppl.\  }
{\bf 190}, 197 (2011)  [{\ftn\tt arXiv:1106.0391}]; D.M.~Regan,
{\ftn\tt  arXiv:1112.5899} and references therein.


\bibitem{lectures} G. Lazarides, {\sl Lect. Notes Phys. }{\bf 592}, 351 (2002)
[{\ftn\tt hep-ph/01 11328}]; G. Lazarides, {\sl J. Phys. Conf.
Ser.} {\bf 53}, 528 (2006) [{\ftn\tt hep-ph/0607032}].

\bibitem{review} D.H.~Lyth and A.~Riotto, {\sl Phys.\
Rept.} {\bf 314}, 1 (1999) [{\tt hep-ph/9807278}]; A.~Mazumdar and
J.~Rocher, {\sl Phys.\ Rept.} {\bf 497}, 85 (2011)
[\arxiv{1001.0993}].

\bibitem{bjorn} S. Clesse, {\sl Phys. Rev. D} {\bf 83}, 063518 (2011)
[\arxiv{1006. 4522}]; H. Kodama, K. Kohri, and K. Nakayama, {\sl
Prog. Theor. Phys.} {\bf 126}, 331 (2011) [\arxiv{1102.5612}];
S.~Clesse and B.~Garbrecht, {\sl Phys.\ Rev.\ D} {\bf 86}, 023525
(2012) [\arxiv{12 04.3540}].



\bibitem{jp} J.~Rocher and M.~Sakellariadou, {\sl J. Cosmol.
Astropart. Phys. }{\bf 03}, 004 (2005) [{\ftn\tt hep-ph/0406120}];
R.~Jeannerot and M.~Postma, \jhep{\bf 05}{2005}{071} [{\ftn\tt
hep-ph/0503146}].

\bibitem{gmb} K.~Nakayama \etal, \jcap{12}{2010}{010} [\arxiv{1007.5152}].


\bibitem{buch} W. Buchm\"uller, V. Domcke and K. Schmitz, \npb{862}{2012}{587}
[\arxiv{1202.6679}].

\bibitem{markjones} A.~Basboll, M.~Hindmarsh and D.R.T.~Jones, \jhep{06}{2011}{115}
[\arxiv{1101.5622}].

\bibitem{mark} R.~Battye, B.~Garbrecht and A.~Moss,
{\sl Phys.\ Rev.\ D} {\bf 81}, 123512 (2010) [\arxiv{1001.0769}];
J.~Urrestilla, N.~Bevis, M.~Hindmarsh and M.~Kunz,
\jcap{12}{2011}{021} [\arxiv{1108.2730}].

\bibitem{lofti} L. Boubekeur and D. Lyth, {\sl J. Cosmol. Astropart.
Phys.} {\bf 07}, 010 (2005) [\hepph{0502047}]; K.~Kohri, C.M. Lin
and D.H. Lyth, \jcap{12}{2007}{004} [\arxiv{0707.3826}]; C.M.~Lin
and K.~Cheung, \jcap{03}{2009}{012} [\arxiv{0812.2731}].



\bibitem{sanidas}  S.A.~Sanidas, R.A.~Battye and B.~W.~Stappers,
{\sl Phys.\ Rev.\ D }{\bf 85}, 122003 (2012)  [{\ftn\tt arXiv:1201.2419}].


\bibitem{pta} R. van Haasteren \etal, \arxiv{1103.0576}.


\bibitem{quin} C. Pallis, \npb{751}{2006}{129} [\hepph{0510234}].


\bibitem{rsym} G. Lazarides and Q. Shafi, {\sl Phys.
Rev. D }{\bf 58}, {071702} ({1998}) [\hepph{9803397}].


\bibitem{Idecay} M.~Endo, F.~Takahashi and T.T.~Yanagida, \prd{76}{2007}{083509} [{\ftn \tt
arXiv:0706.0986}].



\bibitem{brand} M.~Bolz, A.~Brandenburg and W.~{Buchm\"uller},
{\sl Nucl. Phys. }\textbf{B606}, 518 (2001);  M.~Bolz,
A.~Brandenburg and W.~{Buchm\"uller}, \npb{790}{2008}{336} (E)
[\hepph{0012052}]; J.~Pradler and F.D. Steffen,
\prd{75}{2007}{023509} [\hepph{0608344}].


\bibitem{nMCI} C. Pallis and Q. Shafi, \prd{86}{2012}{023523} [{\ftn\tt arXiv: 1204.0252}].



\bibitem{running} S.~Antusch, J.~Kersten, M.~Lindner and M.~Ratz,
\npb{674}{2003}{401} [\hepph{0305273}].




\bibitem{lsenoguz} V.N.~\c{S}eno\u{g}uz,
\prd{76}{2007}{013005} [{\tt\ftn arXiv:07 04.3048}].




\end{thebibliography}
\end{document}